\documentclass{PoS}
\usepackage{lineno}
\usepackage{amsmath}

\newcommand{\sqrtsNN}{\sqrt{s_{\rm \scriptscriptstyle NN}}}

\newcommand{\GeV}{\mathrm{GeV}}
\newcommand{\TeV}{\mathrm{TeV}}
\newcommand{\mub}{\mathrm{\mu b}}
\newcommand{\PbPb}{\mbox{Pb--Pb}}

\newcommand{\Raa}{R_{\rm AA}}
\newcommand{\Taa}{T_{\rm AA}}
\newcommand{\pt}{p_{\rm T}}
\newcommand{\DtoKpi}{{\rm D}^0 \to {\rm K}^-\pi^+}
\newcommand{\DtoKpipi}{{\rm D}^+\to {\rm K}^-\pi^+\pi^+}
\newcommand{\DstartoDpitoKpipi}{{\rm D}^{*+} \to {\rm D}^0 \pi^+ \to {\rm K}^- \pi^+ \pi^+}

\newcommand{\DstophipitoKKpi}{{\rm D_s^{+}\to \phi\pi^+\to K^-K^+\pi^+}}

\newcommand{\Dzero}{{\rm D^0}}
\newcommand{\Dzerobar}{{\overline{{\rm D}}\,^0}}
\newcommand{\Dstar}{{\rm D^{*+}}}

\newcommand{\Dplus}{{\rm D^+}}

\newcommand{\Ds}{{\rm D_s^+}}

\newcommand{\de}{{\rm d}}
\newcommand{\voneodd}{v_1^{\rm odd}}

\title{Measurement of D-meson nuclear modification factor and flow in Pb--Pb collisions with ALICE at the LHC}

\ShortTitle{Measurement of D-meson $\Raa$ and flow in Pb--Pb collisions with ALICE at the LHC}

\author{\speaker{Fabrizio Grosa}\thanks{on behalf of the ALICE Collaboration}\\
        Politecnico di Torino, Corso Duca degli Abruzzi 24, 10129 Torino Italy\\
        INFN sez. Torino, via Pietro Giuria 1, 10125 Torino Italy\\
        E-mail: \email{fabrizio.grosa@cern.ch}}

\abstract{Heavy quarks are sensitive probes of the colour-deconfined medium formed in ultra-relativistic heavy-ion collisions, the Quark-Gluon Plasma (QGP). The ALICE Collaboration measured the production of $\Dzero$, $\Dplus$, $\Dstar$ and $\Ds$ mesons in $\PbPb$ collisions at $\sqrtsNN=5.02~\TeV$. The properties of the in-medium energy loss and the possible modification of the charm-quark hadronisation mechanism are investigated via the measurement of the nuclear modification factor ($\Raa$) of non-strange and strange D mesons. In mid-central collisions, the measurement of the D-meson elliptic flow ($v_2$) at low and intermediate transverse momentum ($\pt$) gives insight into the participation of the charm quark in the collective motion of the system, while at high $\pt$ it constrains the path-length dependence of the energy loss. The coupling of the charm quark to the light quarks in the underlying medium is further investigated with the application of the event-shape engineering (ESE) technique to D-meson elliptic flow. Finally, the role of the early magnetic field created in heavy-ion collisions is studied via the first measurement at the LHC energies of the charged-dependent directed flow ($v_1$) of $\Dzero$ mesons as a function of pseudorapidity.
}

\FullConference{International Conference on Hard and Electromagnetic Probes of High-Energy Nuclear Collisions\\
		30 September - 5 October 2018\\
		Aix-Les-Bains, Savoie, France}

\begin{document}

\section{Introduction}
In ultra-relativistic heavy-ion collisions, heavy quarks (charm and beauty) are produced in the early times of the reaction via hard-scattering processes and they subsequently probe the colour-deconfined medium, known as the Quark-Gluon Plasma (QGP). The measurement of the production of hadrons containing heavy quarks in nucleus--nucleus collisions is used to study the properties of the in-medium energy loss. The comparison to light-flavour hadrons provides information about the quark-mass and colour-charge dependence, while the possible modification of the hadronisation mechanism in the medium can be investigated via the comparison of heavy-flavour hadrons with and without strange-quark content. Further insights into the interaction of heavy quarks with the QGP is given by the measurement of azimuthal anisotropies, which are typically characterised in terms of Fourier coefficients, $v_{\rm n}=\langle \cos{\rm n}(\varphi-\Psi_{\rm n})\rangle$, where $\varphi$ is the particle-momentum azimuthal angle, the brackets denote the average over all the measured particles in the considered events, and $\Psi_{\rm n}$ is the symmetry-plane angle relative to the n$^{th}$ harmonic. The second-harmonic coefficient, called elliptic flow, is the dominant term in mid-central heavy-ion collisions and, at low $\pt$ is sensitive to the participation of the heavy quarks into the collective dynamics of the underlying medium, while at high $\pt$ constraints the path-length dependence of the parton energy loss in the medium. In addition, due to the early production of the charm quarks and their relaxation time, which is similar to the QGP lifetime, the first harmonic coefficient (directed flow) of charmed hadrons was suggested to be a potential probe of the strong initial magnetic field created in heavy-ion collisions that induces electromagnetic currents in the QGP~\cite{Das:2016cwd}. The expected consequence of these currents is a contribution to the pseudorapidity-odd component of directed flow with opposite sign for c and $\rm{\bar{c}}$ quarks, due to the opposite electric charge.

Charmed mesons were reconstructed in ALICE at mid rapidity ($|y|<0.8$) via the decay channels $\DtoKpi$,  $\DtoKpipi$, $\DstartoDpitoKpipi$, and $\DstophipitoKKpi$ and their charge conjugates. Geometrical selections on the decay-vertex topology and particle identification of the decay products were applied to reduce the combinatorial background. The raw D-meson yields were extracted via an invariant-mass analysis. The efficiency and acceptance corrections were obtained from MC simulations based on HIJING \cite{Wang:1991hta} and PYTHIA 6 \cite{Sjostrand:2006za} event generators. The fraction of prompt D mesons was estimated with a FONLL-based approach \cite{Cacciari:1998it,Acharya:2017qps}.

In this contribution, the most recent results on the production and azimuthal anisotropy of D mesons, measured in $\PbPb$ collisions at $\sqrtsNN=5.02~\TeV$, are presented. The data sample, collected in 2015, consisted of about $10^8$ minimum-bias collisions, corresponding to an integrated luminosity of about 13 $\mub^{-1}$.

\section{D-meson nuclear modification factor $\pmb{\Raa}$}
\begin{figure}[t]
\centering
\includegraphics[height=7 cm]{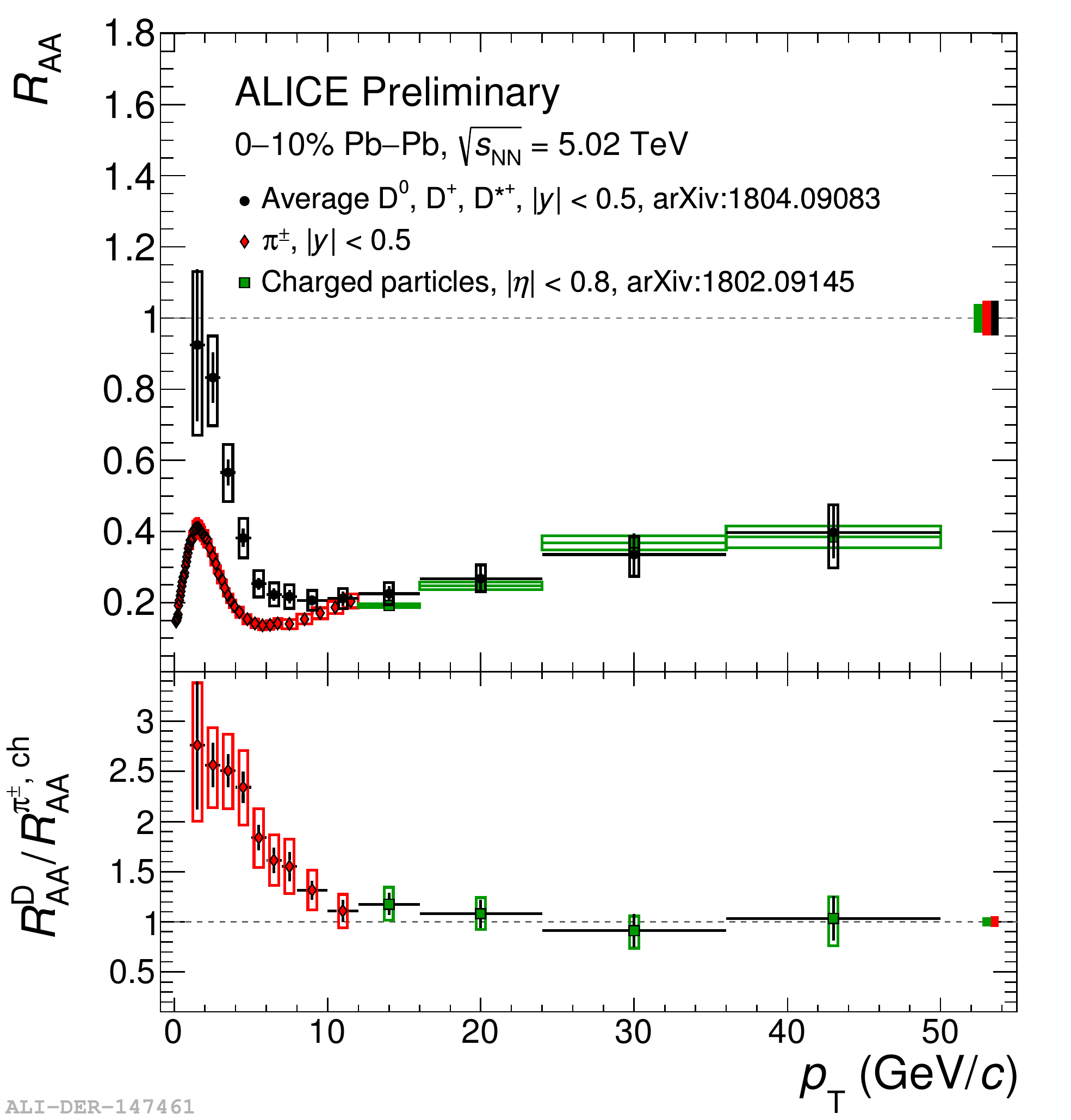}
\hspace{0.5 cm}
\includegraphics[height=7 cm]{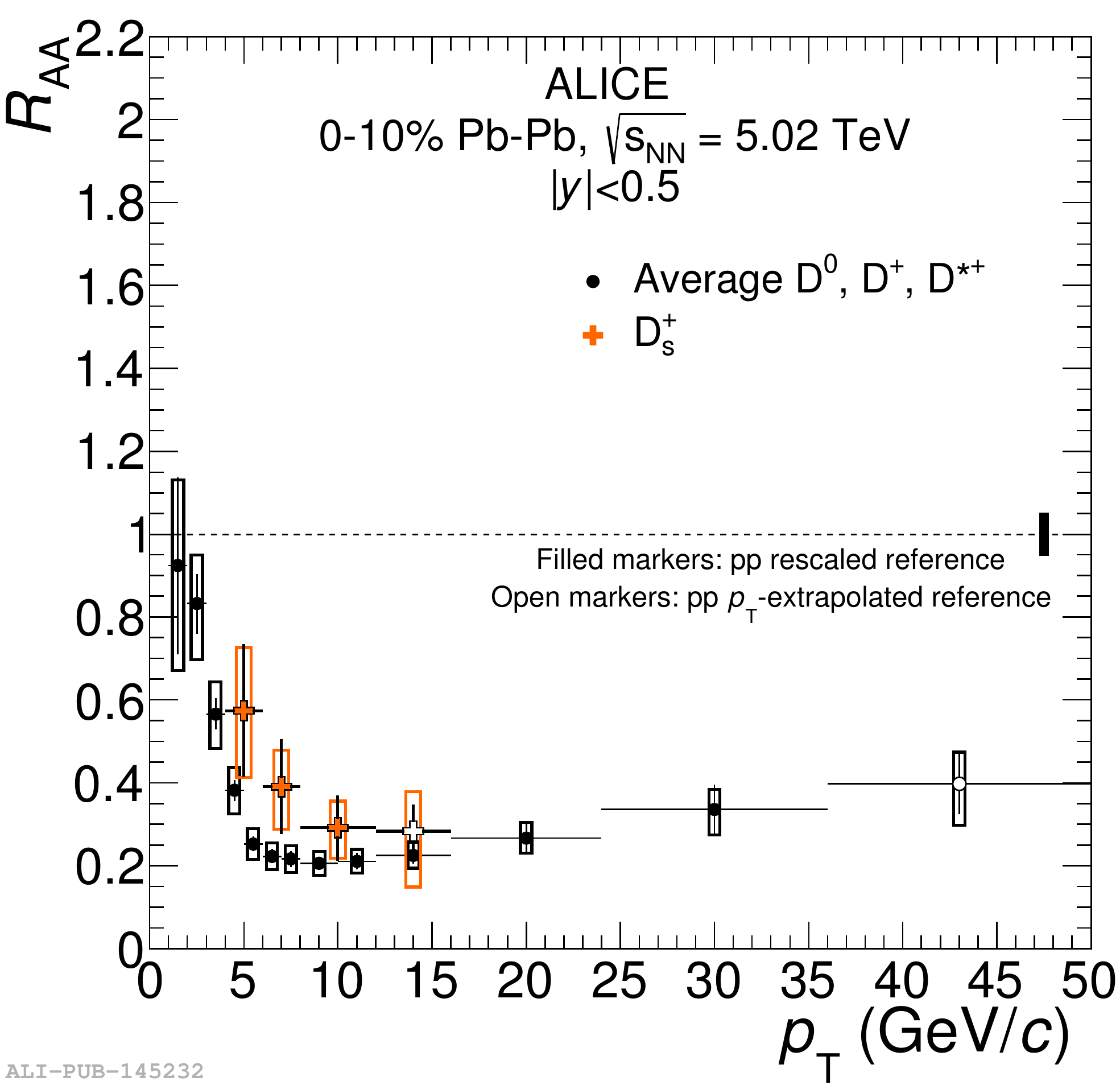}
\caption{Average prompt $\Dzero$, $\Dplus$, and $\Dstar$ $\Raa$ in the 0--10\% centrality class compared to the $\Raa$ of $\pi^\pm$ for $\pt<12~\GeV/c$ and charged particle $\Raa$ for $\pt>12~\GeV/c$ (left), and the $\Raa$ of prompt $\Ds$ (right).}
\label{fig:DmesonRaa}      
\end{figure}

The production of prompt $\Dzero$, $\Dplus$, $\Dstar$, and $\Ds$ mesons was measured in central (0--10\%), mid-central (30--50\%) and peripheral (60--80\%) $\PbPb$ collisions, showing an increasing suppression from peripheral to central collisions compared to pp collisions~\cite{Acharya:2018hre}, quantified by the nuclear modification factor $\Raa = (\de N_{\rm AA}/\de\pt)/(\langle \Taa\rangle\de\sigma_{\rm pp}/\de\pt)$, where $\de N_{\rm AA}/\de\pt$ and $\de\sigma_{\rm pp}/\de\pt$ are the $\pt$-differential yields and cross section in nucleus--nucleus and pp collisions, respectively, while $\langle \Taa\rangle$ is the average nuclear overlap function. The average $\Raa$ of prompt $\Dzero$, $\Dplus$, and $\Dstar$ mesons in the 10\% most central collisions is shown in Fig.~\ref{fig:DmesonRaa}. In the left panel it is compared to the charged-pion (charged-particle~\cite{Acharya:2018qsh}) $\Raa$ for $\pt<12~\GeV/c$ ($\pt>12~\GeV/c$) and in the right panel to the $\Raa$ of prompt $\Ds$ mesons. The $\Raa$ of D mesons is higher than that of charged pions of more than $2\,\sigma$ in each $\pt$ bin for $\pt<8~\GeV/c$, suggesting a dependence on the quark mass of the in-medium energy loss.
This observation is however nontrivial, since several other effects can contribute to this difference, such as the different scaling of soft and hard probes at low $\pt$, the different initial shapes of $\pt$ spectra, radial flow and coalescence. The $\Raa$ values of $\Ds$ mesons are systematically larger than those of non-strange D mesons, as expected in case of a significant contribution of charm-quark hadronisation via coalescence in a strangeness-enhanced medium. However the two measurements are compatible within about one standard deviation of the combined uncertainties.

\section{D-meson elliptic flow $\pmb{v_2}$}
\begin{figure}[t]
\centering
\includegraphics[width=0.8\textwidth]{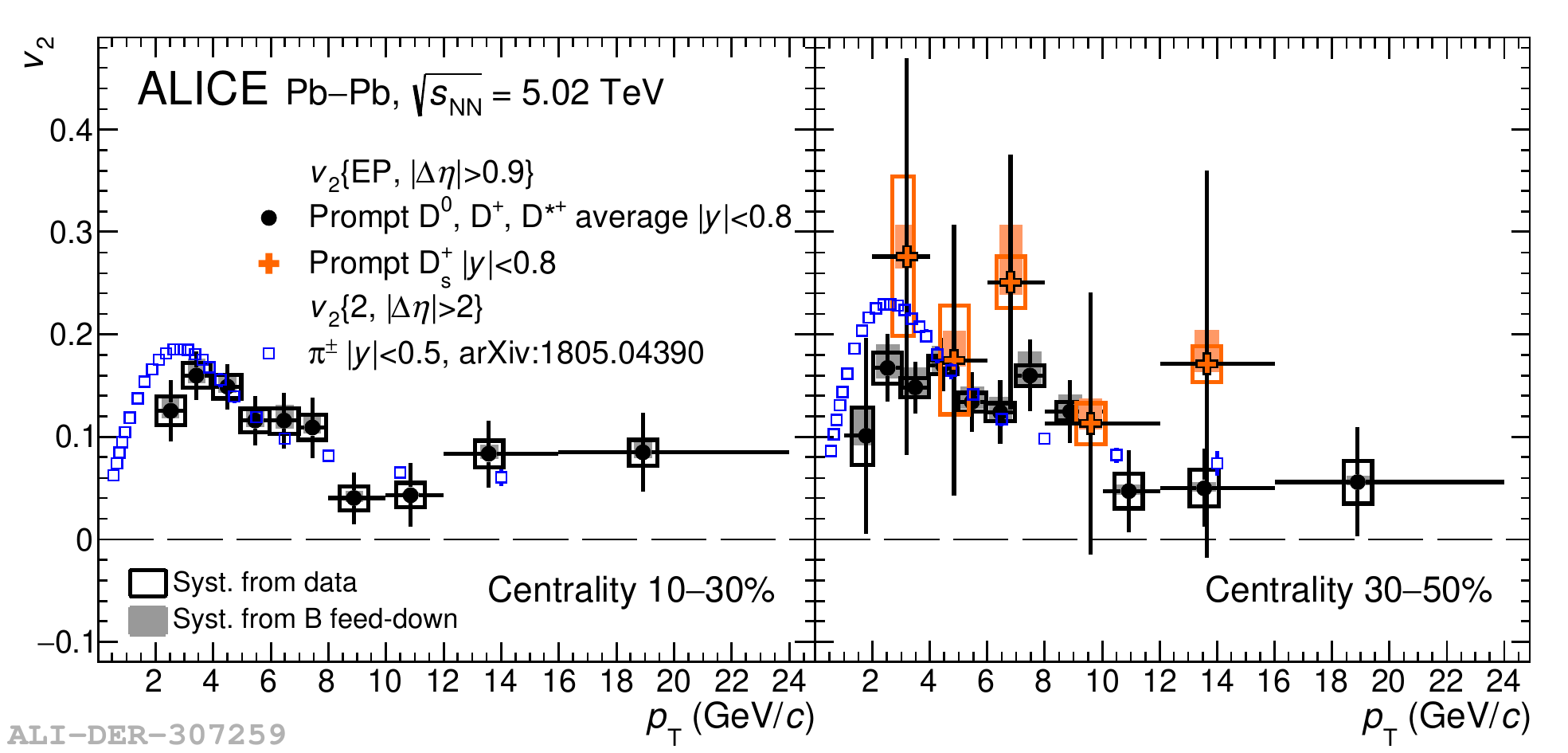}
\caption{Average prompt $\Dzero$, $\Dplus$, and $\Dstar$ $v_2$ in the 10--30\% (left) and 30--50\% (right) centrality classes compared to the $v_2$ of $\pi^\pm$ and that of prompt $\Ds$ in the 30--50\% centrality class.}
\label{fig:Dmesonv2}      
\end{figure}

Figure~\ref{fig:Dmesonv2} shows the average elliptic flow of prompt $\Dzero$, $\Dplus$ and $\Dstar$ measured in mid-central (10--30\% and 30--50\%) $\PbPb$ collisions using the event-plane method~\cite{Acharya:2017qps,Acharya:2018bxo}, compared to that of prompt $\Ds$ in the 30--50\% centrality class and the $v_2$ of charged pions measured with the scalar-product method~\cite{Acharya:2018zuq}. 
The average non-strange D-meson $v_2$ is found to be larger than zero in the range $2<\pt<10~\GeV/c$ and similar to that of charged pions, confirming the participation of the charm quark into the collective expansion of the medium. The $\Ds$ $v_2$ is compatible to that of non-strange D mesons and positive with a significance of about $2.6\,\sigma$. The $v_2$ of $\Dzero$ and $\Dplus$ mesons was further investigated applying the event-shape engineering technique~\cite{Voloshin:2008dg}. This technique relies on the classification of events with fixed centrality but different average elliptic flow, quantified by the magnitude of the second-harmonic reduced flow vector, $q_2=|\vec{Q}|/\sqrt{M}$, where $M$ is the multiplicity and $\vec{Q}_2$ is the second-harmonic flow vector.
The D-meson $v_2$ measured in the 20\% (60\%) of events with largest (smallest) $q_2$ increases (decreases) by about 40\% (25\%), confirming a correlation between the D-meson azimuthal anisotropy and the collective expansion of the bulk matter~\cite{Acharya:2018bxo}.
\section{D-meson directed flow $\pmb{v_1}$}
\begin{figure}[b]
\centering
\includegraphics[width=0.45\textwidth]{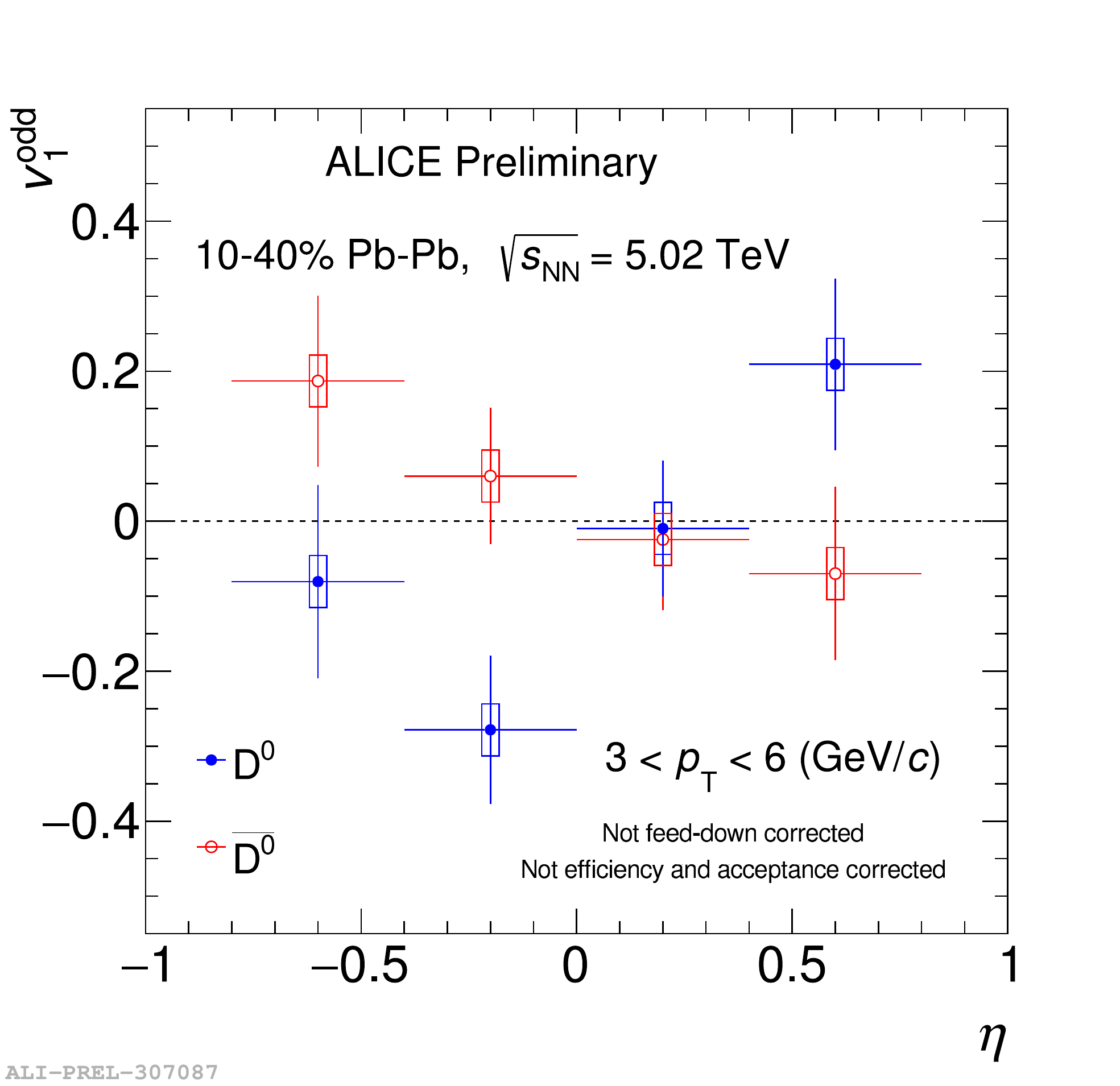}
\includegraphics[width=0.45\textwidth]{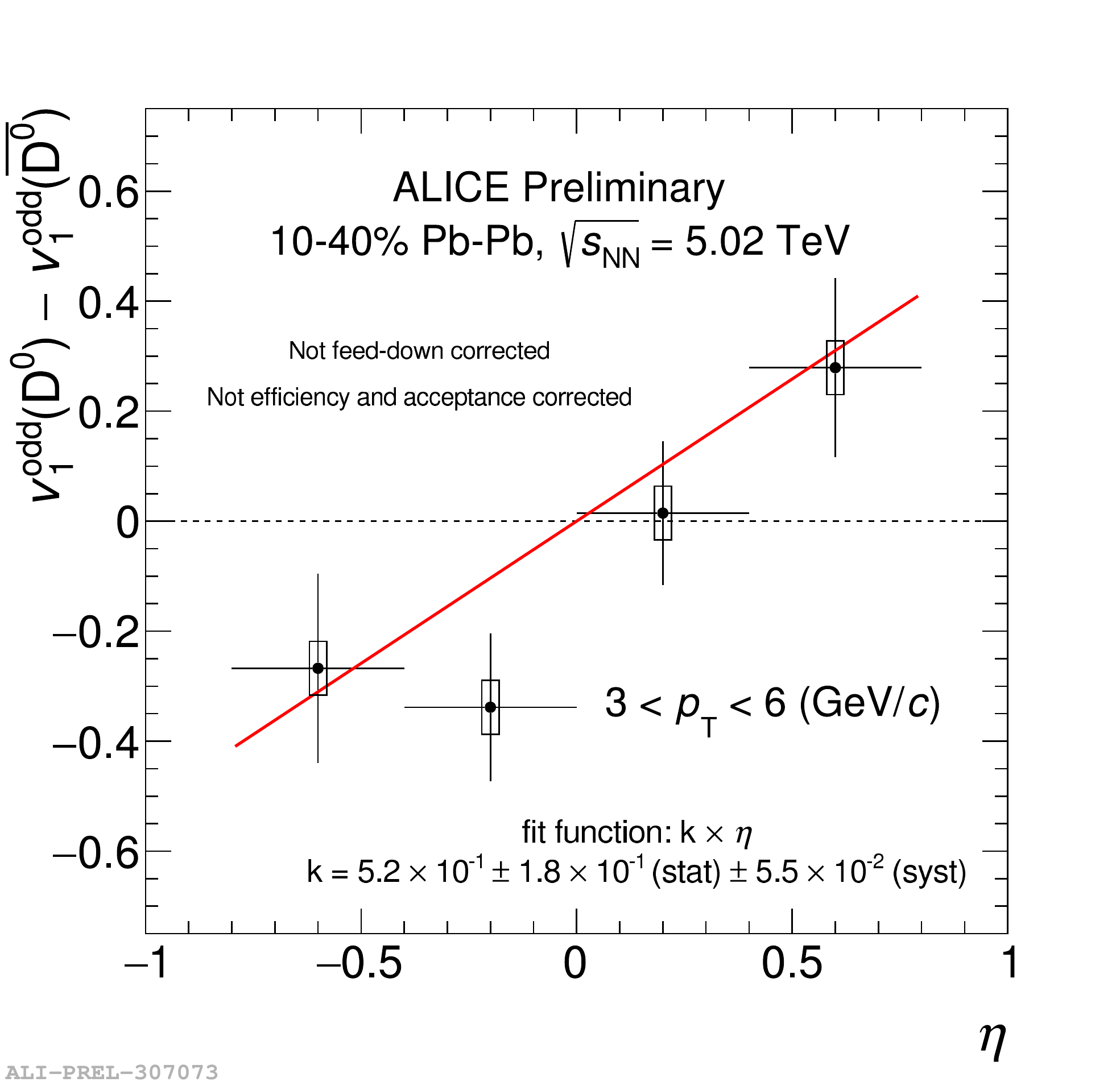}
\caption{$\Dzero$ and $\Dzerobar$ $\voneodd$ (left) and their difference (right) as a function of $\eta$ in the 10--40\% centrality class.}
\label{fig:D0v1}      
\end{figure}
The directed flow of $\Dzero$ and $\Dzerobar$ mesons was measured with the scalar-product method in the 10--40\% centrality class. The spectator plane was reconstructed from the transverse asymmetry of energy deposited by spectator neutrons in two neutron Zero-Degree Calorimeters (ZDCs) located at $\eta>8.8$ (ZDC-A) and $\eta<8.8$ (ZDC-C). The $v_1$ was decomposed into a rapidity-odd component, which is conventionally defined such that the directed flow of the spectator neutrons at positive pseudorapidity has positive sign, $\voneodd = \frac{1}{2}[v_1({\rm ZDC\text{-}A})-v_1({\rm ZDC\text{-}C})]$~\cite{Abelev:2013cva}.  Figure~\ref{fig:D0v1} shows the $\voneodd$ of $\Dzero$ and $\Dzerobar$ with $3<\pt<6~\GeV/c$ (left panel) and their difference, $\Delta\voneodd = \voneodd(\Dzero)-\voneodd(\Dzerobar)$ (right panel), as a function of $\eta$. A hint of signal is observed by fitting the charge difference with a linear function, $\Delta\voneodd=k\times\eta$, from which a positive slope is obtained with $2.7\,\sigma$ significance.

\section{Conclusions}
The ALICE Collaboration has measured the $\Raa$ and the $v_2$ of prompt $\Dzero$, $\Dplus$, $\Dstar$ and $\Ds$ mesons (and charge conjugates) and the charge-dependent $\voneodd$ of $\Dzero$ and $\Dzerobar$ mesons in Pb--Pb collisions at $\sqrtsNN=5.02~\TeV$. 

The D-meson $\Raa$ in central collisions is higher than that of light hadrons for $\pt<8~\GeV/c$. A hint of charm hadronisation via coalescence is provided by the comparison between the $\Raa$ of $\Ds$ and non-strange D mesons. The D-meson $v_2$ is observed to be positive for $2<\pt<10~\GeV/c$ and similar to that of charged pions in mid-central $\PbPb$ collisions.	 Moreover, the application of the event-shape engineering technique shows a positive correlation between the D-meson azimuthal anisotropy and the collective expansion of the bulk of light hadrons. Finally, the first measurement of the $\Dzero$ and $\Dzerobar$ $\voneodd$ at the LHC energies suggests a charge difference, quantified with the slope of $\Delta\voneodd$ as a function of $\eta$, which is found to be positive with a significance of $2.7\,\sigma$.

\end{document}